\documentclass[superscriptaddress,preprint]{revtex4}
\usepackage{mathrsfs}
\usepackage{amssymb}
\usepackage[tbtags]{amsmath}
\usepackage{graphicx}
\usepackage{epsfig,graphicx,times}
\usepackage{color}
\usepackage{subfigure}

\begin{document}

\title{Entangling a series of trapped ions by moving cavity bus}
\author{ZHANG Miao}
\affiliation{Quantum Optoelectronics Laboratory, School of Physics
and Technology, Southwest Jiaotong University, Chengdu 610031,
China}
\author{JIA Huan-Yu}
\affiliation{Quantum Optoelectronics Laboratory, School of Physics
and Technology, Southwest Jiaotong University, Chengdu 610031,
China}
\author{WEI Lian-Fu
\footnote{Email:weilianf@mail.sysu.edu.cn; weilianfu@gmail.com}}
\affiliation{Quantum Optoelectronics Laboratory, School of Physics
and Technology, Southwest Jiaotong University, Chengdu 610031,
China} \affiliation{State Key Laboratory of Optoelectronic Materials
and Technologies, School of Physics and Engineering, Sun Yat-sen
University, Guangzhou 510275, China}
\date{\today}

\begin{abstract}
Entangling multiple qubits is one of the central tasks for quantum
information processings. Here, we propose an approach to entangle a
number of cold ions (individually trapped in a string of microtraps)
by a moved cavity. The cavity is pushed to include the ions one by
one with an uniform velocity, and thus the information stored in
former ions could be transferred to the latter ones by such a moving
cavity bus. Since the positions of the trapped ions are precisely
located, the strengths and durations of the ion-cavity interactions
can be exactly controlled. As a consequence, by properly setting the
relevant parameters typical multi-ion entangled states, e.g., $W$
state for $10$ ions, could be deterministically generated. The
feasibility of the proposal is also discussed.

\end{abstract}

\maketitle

Entanglement is one of the heart concepts in quantum mechanics, and
has potential applications in high precision spectroscopy, quantum
communication, cryptography and computation~\cite{four-particle
entanglement}. Entanglements of two and three particles have been
studied extensively and are well understood. Recently, entanglements
of more than three particles, e.g., four, five, six, seven, and
eight ones, have been successively demonstrated~\cite{four-particle
entanglement,five-particle entanglement, six-particle entanglement,
eight-particle entanglement}. Among the various kinds of entangled
states, the well-known $W$ state plays an important role in quantum
information processes~\cite{QIP}, as its entanglement is maximally
persistent and robust even under particle losses~\cite{W-states}.
Therefore, much interests have been paid recently on the generations
of $W$ states (see, e.g. \cite{eight-particle entanglement}).

To generate an entangled state, the system of trapped ions has many
advantages, such as the convenient manipulations, relative long
coherence time, and easily readout, etc.~\cite{Rev.Mod.trapped-ion}.
Actually, most of the above multi-particle entanglements (e.g., the
eight ions) were first realized experimentally in such a system.
Note that all these experiments were based on either Cirac-Zoller
(CZ)~\cite{Cirac} or M${\text{\o}}$lmer-S${\text{\o}}$rensen
(MS)~\cite{Molmer-Soresen} models. There, a string of ions are
trapped in a single ion-trap, the atomic levels of the trapped ions
act as the qubits and the collective motions of the ions serve as
the data bus.
Assisted by such a data bus the trapped ions could be effectively
entangled under the drivings of the applied laser
beams~\cite{LFWei}.

Here, we propose an alternative approach to entangle a series of
trapped ions by moving cavity bus, instead of the collective motions
of the ions utilized in the famous CZ and MS models. The system we
proposed here is schematized in Fig.~1, wherein cold ions are
individually confined in a string of microtraps~\cite{micro-trap}
and the inter-ion interactions are negligible. A high finesse cavity
is moved to include the ions on the string one by one, and the
cavity mode acts a data bus~\cite{Rev.Mod.cavityQED,data bus} to
transfer the information from one ion to the others.
Comparing to the usual schemes (e.g., the CZ and MS models) using
the moving cavity bus to generate multi-particle entanglement is
conceptually simpler and obviously scalable.

We assume that the velocity of the cavity motion could be properly
set, and thus the cavity could resonantly couple the ions one by one
with the controllable durations.
Also, the ions are assumed to be strongly trapped and well cooled,
such that the regions of the ions' spatial wave packets are much
smaller than the waist and wavelength of the resonant
cavity~\cite{Nature2004,Ion-Cavity-experiments,Ion-Cavity-PRL2002}.
Therefore, the strengths of the cavity-ion couplings are
well-defined (i.e., decided by the located positions of the ions
relative to the cavity mode~\cite{JPB-mode-distribution}).
We shall show that, by properly setting these experimental
parameters, the multi-ion (e.g., from $2$- to $10$-ion) $W$ states
could be prepared within the coherence times of the system.

Without loss of generality, we assume that the interaction between
the moving cavity and the two-level ion is resonant and can be well
described by the well-known Jaynes-Cummings
model~\cite{Rev.Mod.cavityQED}
\begin{equation}
\hat{H}_{\text{JCM}}=\hbar\Omega(x,y,z)(\hat{a}^\dagger\hat{\sigma}_-
+\hat{a}\hat{\sigma}_+).
\end{equation}
Here, $\hat{a}^\dagger$ and $\hat{a}$ are the bosonic creation and
annihilation operators of the cavity field, respectively.
$\hat{\sigma}_+=|e\rangle\langle g|$ and
$\hat{\sigma}_+=|e\rangle\langle g|$ are the raising and lowering
operators of the trapped ion, respectively.
The strength of the ion-field coupling
$\Omega(x,y,z)=\Omega_0f(x,y,z)$ depends on the relative position
$(x,y,z)$ of the trapped ion in the moving cavity, and $\Omega_0$ is
a constant.
Also, the cavity mode is the usual TEM$_{00}$ mode with the
distribution form~\cite{JPB-mode-distribution}
\begin{equation}
f(x,y,z)=e^{-(x^2+y^2)/w^2} \cos(2\pi z/\lambda),
\end{equation}
where $w$ and $\lambda$ are the waist and wavelength of the cavity
mode, respectively.

\begin{figure}[tbp]
\includegraphics[width=9.8cm]{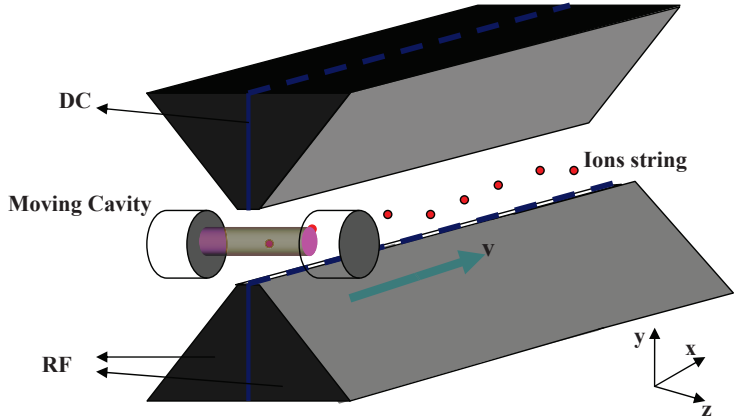}
\caption{(Color online) A sketch of the moving cavity bus to
entangle a series of trapped ions, which are independently trapped
in a string of microtraps.}
\end{figure}

Initially, the system is prepared in the state
$|\varphi_0\rangle=|g_1g_2g_3\cdots g_n\rangle|1\rangle$, i.e., all
the trapped ions (where the index $n$ represents $n$th ion) are in
their atomic ground states and the cavity is in its single-photon
excited state.
Now, we push the cavity (with an uniform velocity $v$ along the $x$
axis) to resonantly excite the trapped ions one by one. In this
process, each ion interacts with the cavity field with an effective
duration $t_{\text{eff}}=\sqrt{\pi}w/v$, and the cavity-ion coupling
strength reads $\Omega=\Omega_0f(0,y,z)$. This is similar to that of
the flying Rydberg atoms interacting with the
cavity~\cite{Rev.Mod.cavityQED}.
Therefore, in this process the system undergoes the following a
series of evolutions:

Firstly, after the cavity excites the first trapped ion (with the
cavity-ion interacting strength $\Omega_1=\Omega_0f(0,y_1,z_1)$),
the system is evolved to the state
\begin{equation}
|\varphi_1\rangle=\cos(\Omega_1t_{\text{eff}})|g_1g_2g_3\cdots
g_n\rangle|1\rangle-i\sin(\Omega_1t_{\text{eff}})|e_1g_2g_3\cdots
g_n\rangle|0\rangle.
\end{equation}

Secondly, after the cavity excites the second trapped ion (with the
coupling strength $\Omega_2=\Omega_0f(0,y_2,z_2)$), the state of
system is
\begin{equation}
\begin{array}{l}
|\varphi_2\rangle=\cos(\Omega_1t_{\text{eff}})\cos(\Omega_2t_{\text{eff}})
|g_1g_2g_3\cdots g_n\rangle|1\rangle\\
\,\,\,\,\,\,\,\,\,\,\,\,\,\,\,\,\,\,
 -i\cos(\Omega_1t_{\text{eff}})
\sin(\Omega_2t_{\text{eff}})|g_1e_2g_3\cdots
g_n\rangle|0\rangle-i\sin(\Omega_1t_{\text{eff}})|e_1g_2g_3\cdots
g_n\rangle|0\rangle.
\end{array}
\end{equation}

Successively, after the $n$th trapped ion is excited (with the
strength $\Omega_n=\Omega_0f(0,y_n,z_n)$) by the moving cavity, we
have a $n$-ion entangled state
\begin{equation}
\begin{array}{l}
|\varphi_n\rangle=-i\big(A_1|e_1g_2g_3\cdots
g_n\rangle+A_2|g_1e_2g_3\cdots g_n\rangle + A_3|g_1g_2e_3\cdots
g_n\rangle
\\
\,\,\,\,\,\,\,\,\,\,\,\,\, \,\,\,\,\,\, +\cdots\cdots
+A_n|g_1g_2g_3\cdots e_n\rangle\big)|0\rangle + B|g_1g_2g_3\cdots
g_n\rangle|1\rangle
\end{array}
\end{equation}
with the amplitudes
\begin{eqnarray}
A_n=\left\{
\begin{array}{l}
\sin(\Omega_1t_{\text{eff}})\,\,\,\,\,\,\,\,\,\,\,\,
\,\,\,\,\,\,\,\,\,\,\,\,\,\,\,\,\,\,\,\,\,\,\,\,\,\,\,
\,\,\,\,\,\,\,\,\,\,\,\,\text{for}\,\,\,\,\,\,n=1,\\
\sin(\Omega_nt_{\text{eff}})\prod_{j=1}^{n-1}\cos(\Omega_jt_{\text{eff}})
\,\,\,\,\,\,\,\,\,\text{for}\,\,\,\,\,\,n\geq2,
\end{array}
\right.
\end{eqnarray}
and $B=\prod_{j=1}^{n}\cos(\Omega_jt_{\text{eff}})$.
Obviously, for preparing the $N$-ion $W$ state, the above
coefficients should be set as $A_n=1/\sqrt{N}$ and $B=0$. This
requires that the coupling strength $\Omega_n$ of $n$th ion
interacting with the cavity should satisfy the condition:
\begin{eqnarray}
\Omega_n=\Omega_0f(0,y_n,z_n)=\frac{v}{w\sqrt{\pi}}\left\{
\begin{array}{l}
\arcsin(\frac{1}{\sqrt{n}})\,\,\,\,\,\,\,\,\,
\,\,\,\,\,\,\,\,\,\,\,\,\,\,\,\,\,\,\,\,\,\,\,\,\,\,\,\,\,\,\,
\,\,\,\,\text{for}\,\,\,\,\,\,n=1,\\
\arcsin[\tan(\Omega_{n-1}\sqrt{\pi}\,w/v)]\,\,\,\,\,\,\,\,
\text{for}\,\,\,\,\,\,n\geq2.
\end{array}
\right.
\end{eqnarray}
This equation can be easily solved by numerical method. Table 1
shows some numerical results for generating the desirable $W$ states
with $\Omega_0\sim2\pi\times14.8$~MHz,
$w\sim10$~$\mu$m~\cite{Coupling1,Coupling2}, $v=800$~m/s and $z=0$.
Above, the ions are assumed to be confined in different position
($y_n$, $z=0$) for achieving different ion-cavity coupling strength
$\Omega_n$. Therefore, the ions should be precisely trapped in any
proper positions, such that $\Omega_n$ are exactly controllable.
Also, the velocity of the cavity should be set sufficiently precise
to achieve the $W$ states with high fidelities.

\begin{table}[h]
\caption{Numerical solutions of $y_n$ ($\mu$m) for generating
multi-ion $W$ states (from $2$ to $10$ ions) with
$\Omega_0=2\pi\times14.8$~MHz, $w=10$~$\mu$m, $v=800$~m/s, and
$z=0$. Note that $y_n$ denotes the relative position of the $n$th
ion in the y-axis.}
\begin{tabular}{cccccccccccc}
\multicolumn{5}{c}{}\\
\hline $N$\,& $y_1$ \,& $y_2$ \,& $y_3$ \,& $y_4$ \,& $y_5$ \,&
$y_6$ \,& $y_7$ \,& $y_8$ \,& $y_9$ \,& $y_{10}$
\\
\hline
2 \,& 9.8204 \,& 5.2083 \\
3 \,& 10.9918 \,& 9.8204 \,& 5.2083 \\
4 \,& 11.7042 \,& 10.9918 \,& 9.8204 \,& 5.2083 \\
5 \,& 12.2126 \,& 11.7042 \,& 10.9918 \,& 9.8204 \,& 5.2083 \\
6 \,& 12.6058 \,& 12.2126 \,& 11.7042 \,& 10.9918 \,& 9.8204 \,& 5.2083 \\
7 \,& 12.9253 \,& 12.6058 \,& 12.2126 \,& 11.7042 \,& 10.9918 \,& 9.8204 \,& 5.2083 \\
8 \,& 13.1936 \,& 12.9253 \,& 12.6058 \,& 12.2126 \,& 11.7042 \,&
10.9918
\,& 9.8204 \,& 5.2083 \\
9 \,& 13.4244 \,& 13.1936 \,& 12.9253 \,& 12.6058 \,& 12.2126 \,&
11.7042
\,& 10.9918 \,& 9.8204 \,& 5.2083 \\
10 \,& 13.6265 \,& 13.4244 \,& 13.1936 \,& 12.9253 \,& 12.6058 \,&
12.2126
\,& 11.7042 \,& 10.9918 \,& 9.8204 \,& 5.2083 \\
\hline
\end{tabular}
\end{table}
\begin{figure}[tbp]
\includegraphics[width=8cm]{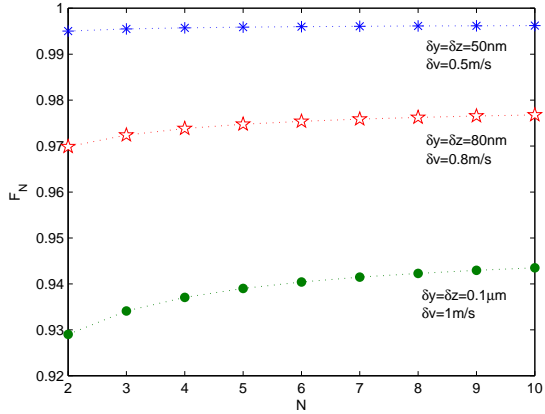}
\caption{The fidelities of the generated multi-ion $W$ states (from
$2$ to $10$ ions).}
\end{figure}

Due to the practically-existing position-uncertain ($\delta y,
\delta z$) of the ion in the cavity and the velocity-fluctuation
$\delta v$ of the cavity motion, the amplitudes $A_n$ of the
generated $W$ states should not be exactly equal to the expected
$1/\sqrt{N}$. Instead, they might read
\begin{eqnarray}
A^{\text{flu}}_n=\left\{
\begin{array}{l}
\sin(\alpha_1)\,\,\,\,\,\,\,\,\,\,\,\,
\,\,\,\,\,\,\,\,\,\,\,\,\,\,\,\,\,\,\,\,
\,\,\,\,\,\,\,\,\,\,\,\,\text{for}\,\,\,\,\,\,n=1,\\
\sin(\alpha_n)\prod_{j=1}^{n-1}\cos(\alpha_j)
\,\,\,\,\,\,\,\,\,\text{for}\,\,\,\,\,\,n\geq2,
\end{array}
\right.
\end{eqnarray}
with
\begin{equation}
\alpha_n=\frac{\Omega_0\sqrt{\pi}\,w}{v\pm\delta
v}\,e^{-(y_n\pm\delta y)^2/w^2} \cos\left[2\pi(z_n\pm\delta
z)/\lambda\right].
\end{equation}
Therefore, the experimentally generated $W_N^{\text{flu}}$ states
might not be the expected $W_N$ states. In order to describe these
departures, a function $F_N=\langle
W_N|W_N^{\text{flu}}\rangle=\sum_{n=1}^NA_nA^{\text{flu}}_n$ could
be introduced to describe the fidelities of the generated $W$
states. Certainly, when $\delta v, \delta y,\delta z\rightarrow0$,
$A_n^{\text{flu}}\rightarrow1/\sqrt{N}$ and thus $F_N\rightarrow1$.

In fact, confining the ions in an exceedingly small region, e.g., on
the order of $10$~nm, has been demonstrated by a number of
experiments (see, e.g.,
\cite{Nature2004,Ion-Cavity-experiments,Ion-Cavity-PRL2002}).
Following the typical experiment reported in Ref.~\cite{Nature2004},
we consider that the cavity is resonant to excite the transition
$^{40}\text{Ca}^+$: $\text{P}_{1/2}\leftrightarrow\text{D}_{3/2}$,
corresponding to a wavelength of $\lambda=866$~nm. Obviously, the
spatial uncertain (e.g., $50$~nm) of the trapped ion is far less
than this wavelength and the waist (with the order of $10$~$\mu$m),
and thus the uncertain of the coupling strength $\Omega_n$ is
significantly small. Consequently, the fidelities of the generated
$W$ states could be high. Fig. 2 shows some numerical results of
fidelities $F_N$ with the typical parameters
$\Omega_0=2\pi\times14.8$~MHz, $w=10$~$\mu$m, $v=800$~m/s, and
$z=0$. Certainly, for the relative smaller fluctuations $\delta
v=0.5$~m/s and $\delta y=\delta z=50$~nm, the fidelities for the
generated multi-ion $W$ states (from $2$ to $10$ ions) can reach to
$99\%$, as shown in Fig. 2.

In addition, the sizes of the experimental microtraps should be
sufficiently small and the cavity should be pushed sufficiently
quick, such that more trapped ions could be entangled within the
finite coherence times of the trapped ions and the cavity. This is
because the total time of the $N$ ions crossing the cavity depends
on the distance $d$ (between the two neighboring trapped ions) and
the velocity $v$ of the moving cavity, as $T=Nd/v$.
For example, the total time to generate the $10$-ion $W$ state can
be less than $0.25$~$\mu$s for $d=20$~$\mu$m and $v=800$~m/s.
Also, it has been experimentally demonstrated~\cite{micro-trap} that
the sizes of the microtraps can be on the order of $10$~$\mu$m by
the photolithography and metal deposition techniques. Thus,
independently trapping the ions with sufficiently small distances,
e.g., $20$~~$\mu$m, is feasible. By improving the experimental setup
to obtain the moving cavity with any modest velocities (e.g.,
$800$~m/s) and velocity-fluctuations (e.g., $0.5$~m/s) are not
exceedingly difficult, in principle.
It has been shown that the spontaneous decay rate on the
$\text{P}_{1/2}\leftrightarrow\text{D}_{3/2}$ is about
$\Gamma=2\pi\times1.69$~MHz~\cite{Nature2004}, and the decay rate of
the cavity (with the finesse $\mathcal{F}=3.5\times10^4$) is about
$\kappa=2\pi\times102$~KHz~\cite{Ion-Cavity-PRL2002}. Not that the
above time for entangling the ions could be further shorted by
enhancing the coupling strength $\Omega_0$ and moving the cavity
more quick. This implies that the present method to generate
multi-ion entanglement within a finite coherence time is possibly
feasible.

In conclusion, we have proposed an idea to entangle a series of
trapped ions: pushing the cavity bus to resonantly excite the
trapped ions one by one. We have shown that by confining the ions in
proper positions the desirable ion-cavity interactions can be
achieved, and thus $W$ states of multi-ion (up to $10$-ion) could be
generated.
Our numerical estimations showed that the existing fluctuations of
the positions of the trapped ions and the velocity of the moving
cavity do not significantly affect the fidelities of the generated
$W$ states. We believe that the proposed model is scalable and might
provide a new way to implement multi-qubit information processings,
although moving the cavity is possibly a new challenge for the
current technique.

This work is partly supported by the NSFC grant No. 10874142,
90921010, and the grant from the Major State Basic Research
Development Program of China (973 Program) (No. 2010CB923104).


\end{document}